\documentstyle[12pt,epsf]{article}

\renewcommand{\thefootnote}{\fnsymbol{footnote}}

\begin{document}
\topskip 2cm 
\begin{titlepage}

\hspace*{\fill}\parbox[t]{4cm}{EDINBURGH 98/21 \\ MSUHEP-80928
\\ Sept 28, 1998 \\ hep-ph/9810215}

\vspace{2cm}

\begin{center}
{\large\bf Virtual Next-to-Leading Corrections to the Lipatov Vertex}\\
\vspace{1.5cm}
{\large Vittorio Del Duca}\footnote{On leave of absence from
I.N.F.N., Sezione di Torino, Italy.}\\
\vspace{.5cm}
{\sl Particle Physics Theory Group,\,
Dept. of Physics and Astronomy\\ University of Edinburgh,\,
Edinburgh EH9 3JZ, Scotland, UK}\\
\vspace{.5cm}
and\\
\vspace{.5cm}
{\large Carl R. Schmidt} \\
\vspace{.5cm}
{\sl Department of Physics and Astronomy\\
Michigan State University\\
East Lansing, MI 48824, USA}\\
\vspace{1.5cm}
\vfil

\begin{abstract}

We compute the virtual next-to-leading corrections to the Lipatov vertex
in the helicity-amplitude formalism. 
These agree with previous results by Fadin and collaborators,
in the conventional dimensional-regularization scheme.
We discuss the choice of reggeization scale in order to minimize
its impact on the next-to-leading-logarithmic corrections to the 
BFKL equation.

\end{abstract}
\end{center}

\end{titlepage}

\setcounter{footnote}{0}
\renewcommand{\thefootnote}{\arabic{footnote}}

\section{Introduction}
\label{sec:0}

Recently, a long awaited calculation of the next-to-leading-logarithmic
(NLL) corrections to the Balitsky-Fadin-Kuraev-Lipatov (BFKL) equation 
\cite{FKL}-\cite{bal} has been completed \cite{nll}.  The BFKL program is
designed to resum the large logarithms of type $\ln(\hat s/|\hat t|)$ in
semi-hard strong-interaction processes, which are characterized by two large
and disparate kinematic scales: $\hat s$, the squared parton
center-of-mass energy and $|\hat t|$, of the order of the squared 
momentum transfer.  At leading logarithmic (LL) accuracy there can be large
dependence on the exact choice of the transverse momentum scale in this 
resummation, which should be reduced in the NLL calculation.
However, the NLL corrections turn out to be large and negative \cite{nll}, and
when the NLL result is applied to Deep Inelastic 
Scattering (DIS) at small $x$ the phenomenological predictions 
are unreliable \cite{ball}.

It has been suggested that the bad behavior of the NLL BFKL resummation
is due to the presence of double logarithms of the ratio of two 
transverse scales, which are ignored in the resummation \cite{salam}. 
On the other hand, an independent check of the NLL 
resummation \cite{nll} is not available yet.  A partial check
has been performed \cite{ciaf}; however, it relies on the same QCD 
amplitudes \cite{real,fl,virtual,linear,ffk} that constitute the 
building blocks of the original NLL calculation, and it assumes 
them to have been computed correctly. 
We have undertaken an independent calculation of the QCD amplitudes 
relevant to the NLL resummation using the helicity-amplitude formalism
\cite{vdd,dds}. 

The solution to the LL BFKL equation is an off-shell gluon 
Green's function, which represents the gluon propagator 
exchanged in the crossed channel of parton-parton scattering.
The building blocks of this equation 
are the Lipatov vertex \cite{FKL}, which summarizes the emission 
of a gluon along the propagator, and the LL reggeization term \cite{fkl2}, 
which summarizes the virtual corrections to the propagator. 
In NLL approximation, one needs the reggeization term to the 
corresponding accuracy \cite{ffk}, plus the real and virtual corrections 
to the Lipatov vertex.  The real corrections to this vertex are given by 
the emission of two gluons or of a $\bar{q} q$ pair at comparable rapidities
\cite{real,vdd}, while the virtual corrections are given by the Lipatov vertex 
at one-loop accuracy \cite{fl,virtual,linear}. In this paper we compute the 
Lipatov vertex at one-loop accuracy using the helicity-amplitude 
formalism\footnote{In order to obtain the one-loop Lipatov vertex, 
one also needs the helicity-conserving vertices at one-loop accuracy
\cite{fl,dds,ff,bds}, even though
these do not enter directly the calculation of the NLL corrections
to the BFKL kernel.}. 

The outline of this paper is as follows. In section 2 we set up the 
formalism needed for extracting the one-loop Lipatov vertex from the
five-gluon one-loop helicity amplitudes.  In section 3 we present these
helicity amplitudes, as obtained from the work of Bern, Dixon, and 
Kosower \cite{bdk}.
In section 4 we consider these amplitudes in the multi-Regge
kinematics in order to extract the Lipatov vertex to $O(\epsilon^0)$ in
the expansion of the space-time dimension used for regularization of 
singularities.  In section 5 we consider the amplitudes in
the limit that the central emitted gluon becomes soft, so as to extract
the vertex to $O(\epsilon)$ in this region.  This is necessary to get
the correct contribution from the infrared singularity when integrating over
the momentum of this gluon in the squared
amplitude.  Finally, in section 6 we
discuss the choice of reggeization scale, and we present our conclusions.

\section{The five-gluon amplitude at high-energy}
\label{sec:prov}

We are interested in the five-gluon amplitude in the 
{\sl multi-Regge kinematics}, which presumes that the produced gluons are
strongly ordered in rapidity and have comparable transverse momenta:
\begin{equation}
y_{a'} \gg y \gg y_{b'}; \qquad |p_{a'\perp}|\simeq|k_\perp|
\simeq|p_{b'\perp}|\, .\label{mrk}
\end{equation}
In this kinematics
the tree-level amplitude for $g_a\,g_b\to g_{a'}\,g\,g_{b'}$
scattering may be written ~\cite{FKL}
\begin{eqnarray}
M^{tree}_{g g \rightarrow g g g} &=& 
2 {s} \left[i g\, f^{aa'c}\, C_{-\nu_a\nu_{a'}}^{gg(0)}(-p_a,p_{a'})
\right]\, {1\over t_1}\, \label{three}\\ & &\quad
\times\  \left[i g\,f^{cdc'}\, 
C^{g(0)}_{\nu}(q_a,q_b)\right]\, {1\over t_2}\, \left[i g\,
f^{bb'c'}\, C_{-\nu_b\nu_{b'}}^{gg(0)}(-p_b,p_{b'}) \right]\, ,\nonumber
\end{eqnarray}
where all external gluons are taken to be outgoing, 
 $q_{i}$ are the momenta transferred in 
the $t$-channel, {\it i.e.} $q_{a}= -p_a-p_{a'}$ and 
$q_{b}=p_b+p_{b'}$, and $ t_i \simeq - |q_{i\perp}|^2$.
The vertices $g^*\, g \rightarrow g$, with $g^*$ an off-shell gluon,
are given by \cite{FKL,ptlip}
\begin{equation}
C_{-+}^{gg(0)}(-p_a,p_{a'}) = -1 \qquad C_{-+}^{gg(0)}
(-p_b,p_{b'}) = - {p_{b'\perp}^* \over p_{b'\perp}}\, ,\label{centrc}
\end{equation}
and the Lipatov vertex $g^*\, g^* \rightarrow g$ \cite{ptlip,lip} is
\begin{equation}
C^{g(0)}_+(q_a,q_b) = \sqrt{2}\, {q^*_{a\perp} q_{b\perp}\over k_{\perp}}\, 
,\label{lipeq}
\end{equation}
with $p_{\perp}=p_x+ip_y$ the complex transverse momentum.
The $C$-vertices transform
into their complex conjugates under helicity reversal,
$C_{\{\nu\}}^*(\{k\}) = C_{\{-\nu\}}(\{k\})$. The helicity-flip
vertex $C^{(0)}_{++}$ is subleading in the high-energy limit.
For gluon-quark scattering, $g_a\,q_b\to g_{a'}\,g\,q_{b'}$,
or quark-quark scattering,
$q_a\,q_b\to q_{a'}\,g\,q_{b'}$, we only need to 
change the relevant vertices $C^{gg(0)}$ to $C^{\bar q q(0)}$ and
exchange the corresponding structure constants with color matrices 
in the fundamental representation \cite{thuile}.
In eq.~(\ref{three}) the mass-shell condition for the intermediate 
gluon in the multi-Regge kinematics has been used,
\begin{equation}
s = {s_1\, s_2\over |k_{\perp}|^2}\, ,\label{mass}
\end{equation}
with $s\equiv s_{ab},\, s_1\equiv s_{a'1},\, s_2\equiv s_{1b'}$. 

The virtual radiative corrections to eq.~(\ref{three}) in
the LL approximation are obtained, to all orders
in $\alpha_s$, by replacing \cite{FKL,bry}
\begin{equation}
{1\over t_i} \to {1\over t_i} 
\left({s_i\over \tau}\right)^{\alpha(t_i)}\, ,\label{sud}
\end{equation}
in eq.~(\ref{three}), with $\alpha(t)$ related to the loop 
transverse-momentum integration
\begin{equation}
\alpha(t) \equiv g^2 \alpha^{(1)}(t) = \alpha_s\, N_c\, t \int 
{d^2k_{\perp}\over (2\pi)^2}\, {1\over k_{\perp}^2
(q-k)_{\perp}^2}\qquad t = q^2 \simeq - q_{\perp}^2\, ,\label{allv}
\end{equation}
and $\alpha_s = g^2/4\pi$. The reggeization scale $\tau$ is much
smaller than any of the $s$-type invariants, $\tau \ll s, s_1, s_2$,
and it is of the order of the $t$-type invariants, $\tau \simeq t_1,
t_2$.  The precise definition of $\tau$ is immaterial to LL accuracy.
The infrared divergence in eq.~(\ref{allv}) can be regularized in 4
dimensions with an infrared-cutoff mass. Alternatively, 
using dimensional regularization
in $d=4-2\epsilon$ dimensions, the integral
in eq.~(\ref{allv}) is performed in $2-2\epsilon$ dimensions, yielding
\begin{equation}
\alpha(t) = g^2 \alpha^{(1)}(t) = 2 g^2\, N_c\, {1\over\epsilon} 
\left(\mu^2\over -t\right)^{\epsilon} c_{\Gamma}\, ,\label{alph}
\end{equation}
with
\begin{equation}
c_{\Gamma} = {1\over (4\pi)^{2-\epsilon}}\, {\Gamma(1+\epsilon)\,
\Gamma^2(1-\epsilon)\over \Gamma(1-2\epsilon)}\, .\label{cgam}
\end{equation}

In order to go beyond the LL approximation and to compute the one-loop
corrections to the Lipatov vertex, we need a prescription that
allows us to disentangle the virtual corrections to the Lipatov vertex
(\ref{lipeq}) from the corrections to the vertices (\ref{centrc}) and 
the corrections 
that reggeize the gluon (\ref{sud}). Such a prescription is supplied by
the general form of the high-energy scattering amplitude, 
arising from a reggeized gluon in the adjoint representation of 
$SU(N_c)$ passed in the $t_1$- and $t_2$-channels~\cite{fl}. 
Since only the dispersive part of the one-loop amplitude 
contributes to the NLL BFKL kernel, we can use the modified prescription
below\footnote{The general form of the amplitude given by the
exchange of reggeized gluons does not hold for the absorptive part of the
one-loop amplitudes, because other 
color structures occur in the high-energy limit. This has been shown
for the four-gluon one-loop amplitude in ref.~\cite{dds} and it is shown
for the five-gluon one-loop amplitude in appendix C.}.
In the helicity basis of eq.~(\ref{three}) this is given by 
\begin{eqnarray}
\lefteqn{ {\rm Disp}\, M^{aa'dbb'}_{\nu_a\nu_{a'}\nu\nu_{b'}\nu_b} = 
2 {s} \left[i g\, f^{aa'c}\, {\rm Disp}\, 
C_{-\nu_a\nu_{a'}}^{gg}(-p_a,p_{a'})\right]\, {1\over t_1}\, 
\left({s_1\over \tau}\right)^{\alpha(t_1)}} \label{pres}\\
&\times&  \left[i g\,f^{cdc'}\,
{\rm Disp}\, C^g_{\nu}(q_a,q_b)\right]\, {1\over t_2}\, 
\left({s_2\over \tau}\right)^{\alpha(t_2)}\, \left[i g\,
f^{bb'c'}\, {\rm Disp}\, C_{-\nu_b\nu_{b'}}^{gg}(-p_b,p_{b'})
\right]\, ,\nonumber
\end{eqnarray}
where
\begin{eqnarray}
\alpha(t) &=& g^2 \alpha^{(1)}(t) + g^4 \alpha^{(2)}(t) + O(g^6)
\nonumber\\
C^{g} &=& C^{g(0)} + g^2 C^{g(1)} + O(g^4) \label{fullv}\\
C^{gg} &=& C^{gg(0)} + g^2 C^{gg(1)} + O(g^4)\, ,\nonumber
\end{eqnarray}
are the loop expansions for the reggeized gluon, the Lipatov vertex,
and the helicity-conserving vertex, respectively.
In the NLL approximation to the BFKL kernel it is necessary
to compute $\alpha^{(2)}(t)$, $C^{g(1)}$, and $C^{gg(1)}$; however, 
to one loop only $C^{g(1)}$ and $C^{gg(1)}$ appear.
Expanding  eq.~(\ref{pres}) 
to $O(g^5)$ and using eq.~(\ref{three}), we obtain
\begin{eqnarray}
\lefteqn{ {\rm Disp}\,M^{aa'dbb'}_{\nu_a\nu_{a'}\nu\nu_{b'}\nu_b} = 
M_5^{\rm tree}\Biggl\{
1\ +\ g^2 \Biggl[\alpha^{(1)}(t_1) \ln{s_1\over \tau}\ +
\alpha^{(1)}(t_2) \ln{s_2\over \tau}} \nonumber\\ &+&
{{\rm Disp}\,C^{gg(1)}_{-\nu_a\nu_{a'}}(-p_a,p_{a'})\over 
C^{gg(0)}_{-\nu_a\nu_{a'}}(-p_a,p_{a'})}\ +\
{{\rm Disp}\,C^{gg(1)}_{-\nu_b\nu_{b'}}(-p_b,p_{b'})\over
C^{gg(0)}_{-\nu_b\nu_{b'}}(-p_b,p_{b'})} +\ {{\rm Disp}\, 
C^{g(1)}_{\nu}(q_a,q_b)\over C^{g(0)}_{\nu}(q_a,q_b)}\Biggr]\Biggr\}
\, .\label{expa}
\end{eqnarray}
Thus, the NLL corrections to $C^{g(1)}$ can be extracted
from the one-loop $g\,g\to g\,g\,g$ amplitude, by subtracting the
one-loop reggeization (\ref{alph}) and the one-loop corrections to
the helicity-conserving vertex.  The latter has been computed in the 
HV and CDR schemes~\cite{fl,dds} and in the dimensional-reduction 
scheme~\cite{dds} and is given by
\begin{eqnarray}
{{\rm Disp}\, C^{gg(1)}_{-+}(-p,p')\over 
C^{gg(0)}_{-+}(-p,p')} &=& 
  c_{\Gamma} \left\{ \left({\mu^2\over -t}\right)^{\epsilon} 
\left[ N_c\left(-{2\over\epsilon^2} +{1\over\epsilon}\ln{\tau\over-t}
- {32\over 9} - {\delta_R\over 6} 
+ {\pi^2\over 2} \right)\right.\right. \nonumber\\ 
&&\left.\left.\qquad\qquad+ {5\over 9}\ N_f - {\beta_0\over 2\epsilon} 
\right] - {\beta_0\over 2\epsilon}\right\}\, ,\label{vert}
\end{eqnarray}
with $\beta_0 = (11N_c-2N_f)/3$ and the regularization scheme (RS)  
parameter
\begin{equation}
\delta_R = \left\{ \begin{array}{ll} 1 & \mbox{HV or CDR scheme},\\
0 & \mbox{dimensional-reduction scheme}\, . \end{array} \right. \label{cp}
\end{equation}
The last term in eq.~(\ref{vert}) is the modified minimal subtraction 
scheme $\overline{\rm MS}$ ultraviolet counterterm.  Note that 
eq.~(\ref{vert}) differs from the result in Ref.~\cite{dds} by the logarithm
term, because in that paper the reggeization scale had been taken to be 
$\tau=-t$.

\section{The one-loop five-gluon amplitude}
\label{sec:provb}
The color decomposition of a tree-level multigluon
amplitude in a helicity basis is \cite{mp}
\begin{equation}
M_n^{tree} = 2^{n/2}\, g^{n-2}\, \sum_{S_n/Z_n} {\rm tr}(\lambda^{d_{\sigma(1)}} 
\cdots
\lambda^{d_{\sigma(n)}}) \, m_n(p_{\sigma(1)},\nu_{\sigma(1)};...;
p_{\sigma(n)},\nu_{\sigma(n)})\, ,\label{one}
\end{equation}
where $d_1,..., d_n$, and $\nu_1,..., \nu_n$ are
respectively the colors and the
polarizations of the gluons, the $\lambda$'s are the color 
matrices\footnote{Note that 
eq.(\ref{one}) differs by the $2^{n/2}$ factor from the expression
given in ref.\cite{mp}, because we use the standard normalization of
the $\lambda$ matrices, ${\rm tr}(\lambda^a\lambda^b) =
\delta^{ab}/2$.} in the
fundamental representation of SU($N_c$) and the sum is over the noncyclic
permutations $S_n/Z_n$ of the set $[1,...,n]$. We take
all the momenta as outgoing, and consider the {\sl maximally 
helicity-violating}
configurations $(-,-,+,...,+)$ for which the gauge-invariant subamplitudes,
$m_n(p_1,\nu_1; ...; p_n,\nu_n)$, assume the form \cite{mp},
\begin{equation}
m_n(-,-,+,...,+) = {\langle p_i p_j\rangle^4\over
\langle p_1 p_2\rangle \cdots\langle p_{n-1} p_n\rangle 
\langle p_n p_1\rangle}\, ,\label{two}
\end{equation}
where $i$ and $j$ are the gluons of negative helicity. The configurations
$(+,+,-,...,-)$ are then obtained by replacing the $\langle p k\rangle$
products with $\left[k p\right]$ products.  We give the formulae for these
spinor products in appendix A.  Using the high-energy limit
of the spinor products (\ref{hpro}), the tree-level amplitude for 
$g\,g\to g\,g\,g$ scattering may be cast in the form (\ref{three}).

The color decomposition of one-loop multigluon amplitudes is also
known \cite{bk1}. For five gluons it is,
\begin{eqnarray}
\lefteqn{M_5^{1-loop}\ =\ } \label{loop}\\
&&\quad 2^{5/2} g^5 \left[\sum_{S_5/Z_5} \,{\rm tr}
(\lambda^{d_{\sigma(1)}} 
\lambda^{d_{\sigma(2)}} \lambda^{d_{\sigma(3)}} 
\lambda^{d_{\sigma(4)}} \lambda^{d_{\sigma(5)}}) \, 
m_{5:1}(\sigma(1), \sigma(2),
\sigma(3), \sigma(4), \sigma(5)) \right. \nonumber\\
&&+ \left. \sum_{S_5/Z_2\times Z_3} {\rm tr}(\lambda^{d_{\sigma(1)}} 
\lambda^{d_{\sigma(2)}}) {\rm tr}(\lambda^{d_{\sigma(3)}} 
\lambda^{d_{\sigma(4)}}\lambda^{d_{\sigma(5)}})\, m_{5:3}(\sigma(1), 
\sigma(2); \sigma(3), \sigma(4), \sigma(5)) \right]\, ,\nonumber
\end{eqnarray}
where  $\sigma(i)$ is a 
shorthand for $p_{\sigma(i)},\nu_{\sigma(i)}$ in the subamplitudes.
The sums are over the permutations of the five color
indices, up to cyclic permutations within each trace. 
The string-inspired decomposition of the $m_{5:1}$ 
subamplitudes~\cite{bdk} is
\begin{equation}
m_{5:1} = N_c A_5^g + \left(4N_c-N_f\right) A_5^f + 
\left(N_c-N_f\right) A_5^s\, ,\label{stri}
\end{equation}
where $A_5^g$, $A_5^f$, and $A_5^s$ get contributions from
an $N=4$ supersymmetric multiplet, an $N=1$ chiral multiplet,
and a complex scalar, respectively.  Also, we have
\begin{equation}
A_5^x = c_{\Gamma}\, m_5\, (V^x+G^x)\, \qquad\qquad x=g,f,s\,
.\label{dec}
\end{equation}

For the NLL BFKL vertex we need the five-gluon one-loop subamplitudes only
in the helicity configurations which are nonzero at tree level.
We write the functions for the $(1,2,3,4,5)$ color order for the two
relevant helicity configurations below.
The functions obtained from the $N=4$ multiplet, $V^g$ and $G^g$, 
are the same for both helicity configurations~\cite{bdk}, 
\begin{eqnarray}
V^g &=& -{1\over\epsilon^2} \sum_{j=1}^5
\left({\mu^2\over -s_{j,j+1}} \right)^{\epsilon} +
\sum_{j=1}^5 \ln\left({-s_{j,j+1}\over -s_{j+1,j+2}}\right) \,
\ln\left({-s_{j+2,j-2}\over -s_{j-2,j-1}}\right)
+ {5\over 6}\pi^2\, -{\delta_R\over 3}\, ,\nonumber\\
G^g &=& 0\, .\label{vertica}
\end{eqnarray}
The other functions depend on the helicity configuration. We define
\begin{equation}
I_{ijkl} = \left[ ij \right] \langle jk\rangle \left[ kl \right]
\langle li\rangle\, .\label{comp}
\end{equation}
For the $(1^-,2^-,3^+,4^+,5^+)$ helicity configuration
we have~\cite{bdk},
\begin{eqnarray}
V^f &=& -{5\over 2\epsilon} - {1\over 2} \left[\ln\left({\mu^2\over -s_{23}}
\right) + \ln\left({\mu^2\over -s_{51}} \right)\right] - 2\, \nonumber\\
G^f &=& {I_{1234}+I_{1245}\over 2\,s_{12}\, s_{51}}
L_0\left({-s_{23} \over -s_{51}}\right) \label{vertic}\\
G^s &=& -{G^f\over 3} + {I_{1234}\,I_{1245}\,(I_{1234}+I_{1245})
\over 3\,s_{12}^3\, s_{51}^3} L_2\left(-s_{23}\over
-s_{51}\right) \nonumber\\
&+&  {I_{1235}^2\over 3\,s_{12}^2 s_{23} s_{51} }\left(1-{s_{35}
\over s_{12}}\right) + {I_{1234}\,I_{1245}\over 6\,s_{12}^2\,s_{23}\,
s_{51}}\, ,\nonumber
\end{eqnarray}
while for the $(1^-,2^+,3^-,4^+,5^+)$ helicity configuration, 
we have
\begin{eqnarray}
V^f &=& -{5\over 2\epsilon} - {1\over 2} \left[\ln\left({\mu^2\over -s_{34}}
\right) + \ln\left({\mu^2\over -s_{51}} \right)\right] - 2\, \nonumber\\
G^f &=& - {I_{1325}+I_{1342}\over 2s_{13}\, s_{51}}
L_0\left({-s_{34}\over -s_{51}}\right) + {I_{1324}\, I_{1342}\over s_{13}^2\,
s_{51}^2} Ls_1\left({-s_{23}\over -s_{51}}, {-s_{34}\over
-s_{51}}\right) \nonumber\\ &+& {I_{1325}\, I_{1352}\over s_{13}^2\, 
s_{34}^2} Ls_1\left({-s_{12}\over -s_{34}}, 
{-s_{51}\over -s_{34}}\right) \label{vertid}\\
G^s &=& -{I_{1324}^2\, I_{1342}^2\over s_{13}^4\, s_{24}^2\, s_{51}^2}
\left[ 2 Ls_1\left({-s_{23}\over -s_{51}}, {-s_{34}\over
-s_{51}}\right) + L_1\left({-s_{23}\over -s_{51}}\right)
+ L_1\left({-s_{34}\over -s_{51}}\right)\right] \nonumber\\
&& - {I_{1325}^2\, I_{1352}^2\over s_{13}^4\, s_{25}^2\, s_{34}^2}
\left[2 Ls_1\left({-s_{12}\over -s_{34}}, {-s_{51}\over -s_{34}}\right)
+ L_1\left({-s_{12}\over -s_{34}}\right) + L_1\left({-s_{51}\over 
-s_{34}}\right) \right] \nonumber\\ &+&
{2\over 3} {I_{1324}^3\, I_{1342}\over s_{13}^4\, s_{24}\, s_{51}^3}
L_2\left(-s_{23}\over -s_{51}\right) +
{2\over 3} {I_{1352}^3\, I_{1325}\over s_{13}^4\, s_{25}\, s_{34}^3}
L_2\left(-s_{12}\over -s_{34}\right) \nonumber\\
&+& {1\over 3 s_{51}^3}\, L_2\left(-s_{34}\over -s_{51}\right)
\left[ - {I_{1325}\,I_{1342}\, (I_{1325}+I_{1342})\over s_{13}^3}
+ 2 {I_{1342}^3\, I_{1324}\over s_{13}^4\, s_{24}} +
2 {I_{1325}^3\, I_{1352}\over s_{13}^4\, s_{25}} \right] \nonumber\\
&+& {I_{1325}+I_{1342} \over 6\,s_{13}\, s_{51}} 
L_0\left({-s_{34} \over -s_{51}}\right) + {I_{1325}^2\, I_{1342}^2\,
\over 3\, s_{13}^4\, s_{23}\, s_{51}\, s_{34}\, s_{12} } \nonumber\\
&+& {I_{1324}^2\, I_{1342}^2 \over 3 \, s_{13}^4\,s_{23}\,s_{24}\,
s_{34}\, s_{51}} + {I_{1325}^2\, I_{1352}^2 \over 3\, 
s_{13}^4\,s_{25}\,s_{12}\, s_{34}\, s_{51}} 
- {I_{1342}\, I_{1325}\over 6\, s_{13}^2
s_{34}\, s_{51}}\, ,\nonumber
\end{eqnarray}
with the functions $L_0, L_1, L_2, Ls_1$ defined in Appendix B.
For both the helicity configurations above, the functions $V^s$ and
$V^f$ are related by,
\begin{equation}
V^s = -{V^f\over 3} + {2\over 9}\, .\label{rel}
\end{equation}
In the expansion in $\epsilon$, eq.~(\ref{vertica}-\ref{rel}) are 
valid to $O(\epsilon^0)$.
The amplitude~(\ref{loop}) defined in terms of
eq.~(\ref{stri}-\ref{rel}) is $\overline{\rm MS}$ regulated.
Using eq.~(\ref{dec}, \ref{vertica}, \ref{rel}), we can write
the $m_{5:1}$ subamplitude (\ref{stri}) as the sum of a universal
piece, which is the same for both helicity configurations, and a 
non-universal piece, which depends on the helicity configuration,
\begin{equation}
m_{5:1} = m_{5:1}^u + m_{5:1}^{nu}\ ,\label{decomp}
\end{equation}
with
\begin{eqnarray} 
m_{5:1}^u &=& c_{\Gamma}\, m_5\, N_c\, V^g\, ,\label{uni}\\
m_{5:1}^{nu} &=& c_{\Gamma}\, m_5\, \left[\beta_0 V^f + \left(
4N_c-N_f\right) G^f + \left(N_c-N_f\right) \left(G^s+{2\over
9}\right)\right]\, .\nonumber
\end{eqnarray}
In addition,
\begin{equation}
m_{5:3}(4,5;1,2,3) = {1\over N_c}\, 
\sum_{{\rm COP}_4^{(1,2,3)}} m_{5:1}(\sigma(1), 
\sigma(2); \sigma(3), \sigma(4), 5)\ ,\label{nonp}
\end{equation}
where only the $N_f$-independent, unrenormalized contributions to
$m_{5:1}$ are included~\cite{bdk} and 
${\rm COP}_4^{(1,2,3)}$ denotes the subset of
permutations of ${\rm S}_4$ that leave the ordering of (1,2,3)
unchanged up to a cyclic permutation~\cite{bk1}. 

\section{The one-loop corrections to the Lipatov vertex}
\label{sec:oneloop}

To obtain the next-to-leading logarithmic corrections to the Lipatov
$g^*\, g^*\to g$ vertex, we need the amplitude 
$M_5^{1-loop}(B^-,A^-,A'^+,k^+,B'^+)$ in the high-energy limit.
We must consider each of the color orderings in eq.~(\ref{loop})
and expand the subamplitudes~(\ref{stri}) in
powers of $t/s$, retaining only the leading power, which yields
the leading and next-to-leading terms in $\ln(s/t)$.
In fact, at NLL we only need to keep the dispersive 
parts of the subamplitudes $m_{5:1}$ and $m_{5:3}$. By direct
inspection of eq.~(\ref{dec}-\ref{rel}) it is straightforward
to show that if a given color ordering of $m_{5}$ 
is suppressed by a power of $\tilde{t}/\tilde{s}$ at tree-level, where
$\tilde{t} = t_1,\,t_2,\,|k_{\perp}|^2$ and $\tilde{s} = s,\,s_1,\,s_2$,
then the corresponding color ordering of $m_{5:1}$ will also
be suppressed at one-loop. 

For the $M_n^{tree}$ amplitude in the multi-Regge kinematics,
the leading color orderings are obtained by untwisting the color flow,
in a such a way to obtain a double-sided color-flow diagram,
and by retaining only
the color-flow diagrams which exhibit strong rapidity orderings of 
the gluons on both sides of the diagram, without regard for the
relative rapidity ordering between the two sides~\cite{ptlip,ptmr}.
Easy combinatorics then show that there are $2^{n-2}$ such color 
orderings, and because of the reflection and cyclic symmetries of the
subamplitudes only $2^{n-3}$ need to be determined, e.g. all the ones
which have at least $(n-2)/2$ gluons on one side of the color-flow
diagram. For the $M_5^{tree}$ amplitude, and thus for $m_{5:1}$,
we have eight leading color orderings, out of which four need to be
determined, and we can choose them to be $(A^-,A'^+,k^+,B'^+,B^-)$, 
$(A^-,A'^+,k^+,B^-,B'^+)$, $(A^-,A'^+,B'^+,B^-,k^+)$ and 
$(A^-,k^+,B'^+,B^-,A'^+)$. The other four leading subamplitudes are 
then obtained by taking the ones above in reverse order, which yields
an overall minus sign.

For positive values of the invariants $s_{ij}$ we use the
prescription $\ln(-s_{ij}) = \ln(s_{ij}) -i\pi$.   Thus,
in the multi-Regge kinematics and at NLL the dispersive part of the
universal piece~(\ref{vertica}) becomes, using the spinor 
products~(\ref{hpro}),
\begin{eqnarray}
{\rm Disp}\, V^g &=& 
- {1\over\epsilon^2}\, \left[
2\left({\mu^2\over -t_1}\right)^\epsilon + 2\left({\mu^2\over -t_2}
\right)^\epsilon + \left({\mu^2\over |k_{\perp}|^2}\right)^\epsilon\,\right] 
\nonumber\\ &&+ 
{2\over\epsilon}\, \left[
\left({\mu^2\over -t_1}\right)^\epsilon y_1
 +\left({\mu^2\over -t_2}\right)^\epsilon y_2\right]
-{1\over 2} \ln^2{t_1\over t_2}
- {\delta_R\over 3} + {4\over 3}\pi^2\, ,\label{vg}
\end{eqnarray}
where we have written the leading logarithms in terms of the 
physical rapidity intervals 
$y_1 = \ln(s_1/\sqrt{-t_1}|k_{\perp}|)$ and 
$y_2 = \ln(s_2/\sqrt{-t_2}|k_{\perp}|)$.
The non-universal piece~(\ref{uni}) becomes, after
rewriting all the phases in terms of $q_{a\perp}\,q_{b\perp}^\ast$
and some algebraic manipulation,
\begin{eqnarray}
\lefteqn{ m_{5:1}^{nu}\left(\sigma(B^-),\sigma(A^-),\sigma(A'^+),
\sigma(k^+),\sigma(B'^+)\right) } \nonumber\\ &=& 
m_5\left(\sigma(B^-),\sigma(A^-),\sigma(A'^+),
\sigma(k^+),\sigma(B'^+)\right)\, c_{\Gamma} \nonumber\\
&\times& \left\{ -{3\over 2}{\beta_0\over\epsilon} -
{\beta_0\over 2\epsilon}\, \left[ \left(
{\mu^2\over -t_1}\right)^\epsilon + \left({\mu^2\over
-t_2}\right)^\epsilon \right] - {64\over 9}\, N_c\, +
{10\over 9}\,N_f \right.\nonumber\\
&-& {\beta_0\over 2} \left(t_1 + t_2 + 2 q_{a\perp}
q_{b\perp}^\ast\right) {L_0(t_1/t_2)\over
t_2} \label{nlla}\\ &+& \left. {N_c-N_f\over 3}\, |k_{\perp}|^2\,
\left[ - [ 2t_1 t_2 + (t_1 + t_2 + 2|k_{\perp}|^2) q_{a\perp} 
q_{b\perp}^\ast ] {L_2(t_1/t_2)\over t_2^3} - {q_{a\perp} 
q_{b\perp}^\ast \over 2\, t_1\, t_2} \right]
 \right\}\, ,\nonumber
\end{eqnarray}
where the first term is the $\overline{\rm MS}$ 
ultraviolet counterterm, and where 
the permutations $\sigma$ span the eight leading color orderings.
Combining eq.~(\ref{vg}) and (\ref{nlla}),
we see that the dispersive parts of the leading $m_{5:1}$
subamplitudes are all proportional to the corresponding tree-level
subamplitudes.  Therefore, by the tree-level U(1) decoupling 
equations~\cite{bk1,bg} the $m_{5:3}$ subamplitudes vanish,
\begin{equation}
{\rm Disp}\, m_{5:3}\left(\sigma(B^-),\sigma(A^-),\sigma(A'^+),
\sigma(k^+),\sigma(B'^+)\right) = 0 + O(t/s)\, .\label{non}
\end{equation}
Thus, we conclude that the dispersive part of the one-loop five-gluon
amplitude is simply proportional to the tree amplitude to leading
power in $t/s$.  Combining eq.~(\ref{vg}) and 
(\ref{nlla}), it is given to $O(\epsilon^0)$ by
\begin{eqnarray}
\lefteqn{ {\rm Disp}\, M_5^{1-loop}(A^-,A'^+,k^+,B'^+,B^-) = 
M_5^{tree}(A^-,A'^+,k^+,B'^+,B^-)\, g^2\, c_\Gamma}\nonumber\\
&\times& 
\left\{N_c\, \left[ 
- {1\over\epsilon^2}\, \left[
2\left({\mu^2\over -t_1}\right)^\epsilon + 2\left({\mu^2\over -t_2}
\right)^\epsilon + \left({\mu^2\over |k_{\perp}|^2}\right)^\epsilon\,\right]
\right.\right. 
\nonumber\\ &&\left.+ {2\over\epsilon}\, \left[
\left({\mu^2\over -t_1}\right)^\epsilon y_1
 +\left({\mu^2\over -t_2}\right)^\epsilon y_2\right]
-{1\over 2} \ln^2{t_1\over t_2}
- {\delta_R\over 3} + {4\over 3}\pi^2\right]\nonumber\\
&&-{3\over 2}{\beta_0\over\epsilon} -
{\beta_0\over 2\epsilon}\, \left[ \left(
{\mu^2\over -t_1}\right)^\epsilon + \left({\mu^2\over
-t_2}\right)^\epsilon \right] - {64\over 9}\, N_c\, +
{10\over 9}\,N_f \label{oneloop}\\
&&- {\beta_0\over 2} \left(t_1 + t_2 + 2 q_{a\perp}
q_{b\perp}^\ast\right) {L_0(t_1/t_2)\over
t_2} \nonumber\\ &&+ \left. {N_c-N_f\over 3}\, |k_{\perp}|^2\,
\left[ - [ 2t_1 t_2 + (t_1 + t_2 + 2|k_{\perp}|^2) q_{a\perp} 
q_{b\perp}^\ast ] {L_2(t_1/t_2)\over t_2^3} - {q_{a\perp} 
q_{b\perp}^\ast \over 2\, t_1\, t_2} \right]
 \right\}\, .\nonumber
\end{eqnarray}
Note that only the real part of this amplitude contributes to the NLL
corrections to the BFKL equation.  It can easily be obtained using
${\rm Re}(q_{a\perp}q_{b\perp}^\ast)=-(t_1+t_2+|k_{\perp}|^2)/2$.

Using eq.~(\ref{alph}, \ref{expa}, \ref{vert}) and eq.~(\ref{oneloop}),
 we can extract the NLL corrections to the Lipatov vertex
to $O(\epsilon^0)$
\begin{eqnarray}
\lefteqn{ {{\rm Disp}\, C^{g(1)}_{\nu}(q_a,q_b)\over 
C^{g(0)}_{\nu}(q_a,q_b)} = c_\Gamma\, \Biggl\{N_c\, \Biggl[ {1\over\epsilon}\, 
\left({\mu^2\over -t_1}\right)^\epsilon \ln{\tau\over
|k_{\perp}|^2} + {1\over\epsilon}\,\left({\mu^2\over -t_2}\right)^\epsilon
\ln{\tau\over |k_{\perp}|^2} } \nonumber\\ &-&
{1\over\epsilon^2}\, \left({\mu^2\over |k_{\perp}|^2}\right)^\epsilon
-{1\over 2} \ln^2{t_1\over t_2} + {\pi^2\over 3}\Biggr] 
- {\beta_0\over 2\epsilon} - {\beta_0\over 2} \left(t_1 + t_2 + 2 q_{a\perp}
q_{b\perp}^\ast\right) {L_0(t_1/t_2)\over
t_2} \label{looplip}\\ &+& {N_c-N_f\over 3}\, |k_{\perp}|^2\,
\left[ - [ 2t_1 t_2 + (t_1 + t_2 + 2|k_{\perp}|^2) q_{a\perp} 
q_{b\perp}^\ast ] {L_2(t_1/t_2)\over t_2^3} - {q_{a\perp} 
q_{b\perp}^\ast \over 2\, t_1\, t_2} \right]
 \Biggr\} + O(\epsilon)\, .\nonumber
\end{eqnarray}
Note that the dependence on the RS parameter $\delta_R$
has disappeared. If we take $\tau=\mu^2$, eq.~(\ref{looplip})
agrees with the NLL corrections to the Lipatov vertex computed
in ref.~\cite{fl,virtual} in the CDR scheme, to $O(\epsilon^0)$.

\section{The one-loop Lipatov vertex in the soft limit}
\label{sec:soft}

In the NLL corrections to the BFKL kernel, the one-loop Lipatov vertex
is multiplied by the corresponding  tree-level vertex
with the intermediate gluon $k$ integrated over its
phase space. In the soft limit for the intermediate gluon, $k\to 0$, 
an infrared divergence arises\footnote{
No collinear divergence occurs here due to the strong ordering in 
rapidity.}, which in dimensional regularization
manifests itself as a pole of $O(1/\epsilon)$.
Thus, in order to generate correctly all the finite terms in the squared 
amplitude, we need the one-loop Lipatov vertex to $O(\epsilon)$ in the limit
that the gluon $k$ is soft. This can be obtained by computing the 
five-gluon one-loop amplitude to $O(\epsilon)$ in the soft limit 
and then matching on to our previous $O(\epsilon^0)$ result, 
eq.~(\ref{oneloop}).

We find the soft limit of the five-gluon one-loop amplitude by exploiting 
the factorization of the soft singularities in the color-ordered 
subamplitudes:
\begin{eqnarray}
\lefteqn{ m_n^{\rm 1-loop}(..., A,k^\pm, B,...)|_{k\to 0} =} \\ 
& & {\rm Soft}^{\rm tree}(A,k^\pm, B)\, m_{n-1}^{\rm 1-loop}(..., A, B,...) 
+ {\rm Soft}^{\rm 1-loop}(A,k^\pm, B)\, m_{n-1}^{\rm tree}(..., A, B,...)\ .
\nonumber
\end{eqnarray}
This requires the use of one-loop soft functions and four-gluon one-loop
amplitudes,
which have been evaluated to all orders in $\epsilon$ \cite{bds}.
Using these results, the five-gluon one-loop 
amplitude, in the soft limit for the intermediate gluon, is \cite{bds}
\begin{eqnarray}
\lefteqn{ {\rm Disp}\, M_5^{1-loop}(A^-,A'^+,k^\pm,B'^+,B^-)|_{k\to 0} =  
M_5^{tree}(A^-,A'^+,k^\pm,B'^+,B^-)|_{k\to 0}\, g^2\, c_{\Gamma}\, } 
\nonumber\\ &&\qquad\qquad\times \Biggl\{ \left({\mu^2\over - t}
\right)^{\epsilon}
\Biggl[ N_c\, \Biggl[-{4\over\epsilon^2} + {2\over\epsilon}\,
\left(\psi(1+\epsilon) - 2\psi(1-\epsilon) + \psi(1) + \ln{s\over -t}\right) 
\label{softlim}\\ &&\qquad\qquad\qquad+\ {1\over \epsilon (1-2\epsilon)}\,
\left({1 - 
\delta_R\epsilon\over 3-2\epsilon} - 4\right)\Biggr] + N_f\,
{2(1-\epsilon)\over \epsilon (1-2\epsilon)(3-2\epsilon)}  \Biggr] 
\nonumber\\ &&\qquad\qquad\qquad\ - N_c \left({\mu^2\over 
|k_{\perp}|^2}\right)^{\epsilon}\, {1\over\epsilon^2}\,
\left[1 + \epsilon \psi(1-\epsilon) - \epsilon \psi(1+\epsilon)\right]
-{3\over2}{\beta_0\over\epsilon}
\Biggr\}\, .\nonumber
\end{eqnarray}
Eq.~(\ref{softlim}) is valid to all orders in $\epsilon$ for
$\delta_R=0$ and 1. Expanding it to $O(\epsilon)$, yields
\begin{eqnarray}
\lefteqn{ {\rm Disp}\, M_5^{1-loop}(A^-,A'^+,k^\pm,B'^+,B^-)|_{k\to 0} =  
M_5^{tree}(A^-,A'^+,k^\pm,B'^+,B^-)|_{k\to 0}\, g^2\, c_{\Gamma}\, } 
\nonumber\\ &&\quad\times\ \Biggl\{ \left({\mu^2\over - t}\right)^{\epsilon}
\Biggl[ N_c\, \Biggl(-{4\over\epsilon^2} + {2\over\epsilon}\ln{s\over -t}
+ \pi^2 - {64\over 9} - {\delta_R\over 3} + 2\zeta(3)\epsilon 
- {380\over 27}\epsilon - {8\over 9}\delta_R\epsilon \Biggr) \label{softexp}\\
&&\quad\qquad-\ {\beta_0\over \epsilon} + N_f \left({10\over 9} + {56\over 27}
\epsilon\right) \Biggr] - N_c \left({\mu^2\over 
|k_{\perp}|^2}\right)^{\epsilon}\, \left({1\over\epsilon^2} - 
{\pi^2\over 3}\right)  
-{3\over2}{\beta_0\over\epsilon}
+ O(\epsilon^2)\Bigg\} \, .\nonumber
\end{eqnarray}
Matching eq.~(\ref{softlim}) or eq.~(\ref{softexp}) to the
five-gluon one-loop amplitude to $O(\epsilon^0)$, eq.~(\ref{oneloop}),
one can obtain the five-gluon one-loop amplitude with soft corrections to
all orders in $\epsilon$ or to $O(\epsilon)$, respectively.
However, for our purposes it suffices to perform the matching directly
on the one-loop Lipatov vertex. In order to do that, we need the 
one-loop correction to the helicity-conserving impact factor, which
was given to $O(\epsilon^0)$ in eq.~(\ref{vert}), to all orders
in $\epsilon$. It is~\cite{bds},
\begin{eqnarray}
{{\rm Disp}\, C^{gg(1)}_{-+}(-p,p')\over 
C^{gg(0)}_{-+}(-p,p')} &=&
c_{\Gamma}\, \Biggl\{ \left({\mu^2\over - t}\right)^{\epsilon}
\Biggl[ N_c\, \Biggl[-{2\over\epsilon^2} 
+ {1\over\epsilon}\,
\Bigl(\ln{\tau\over-t}+\psi(1+\epsilon) - 2\psi(1-\epsilon) + \psi(1)\Bigr) 
\nonumber
\\ 
+&&\!\!\!\!\!\!\!\!\!\!\!\!\! {1\over \epsilon (1-2\epsilon)}\,
\left({1 - 
\delta_R\epsilon\over 2(3-2\epsilon)} - 2\right)\Biggr]
+ N_f\,
{(1-\epsilon)\over \epsilon (1-2\epsilon)(3-2\epsilon)}  \Biggr] 
-{\beta_0\over2\epsilon}
\Biggr\}\ .\label{allvert}
\end{eqnarray}
Eq.~(\ref{allvert}) is valid to all orders in $\epsilon$ for
$\delta_R=0$ and 1. Using eq.~(\ref{softlim}) and eq.~(\ref{alph},
\ref{expa}, \ref{allvert}) with $t_1=t_2=t$, and the mass-shell 
condition (\ref{mass}), we extract the 
NLL corrections to the Lipatov vertex in the soft limit of the emitted gluon,
\begin{eqnarray}
\left.{{\rm Disp}\, C^{g(1)}_{\nu}(q_a,q_b)\over C^{g(0)}_{\nu}(q_a,q_b)}
\right|_{k\to 0} &&=\  
c_\Gamma\, \Biggl\{N_c\, \Biggl[ {2\over\epsilon}\, \left({\mu^2\over -t}
\right)^\epsilon\,\ln{\tau\over |k_{\perp}|^2} 
 \label{nllall}\\ &&\qquad-\
{1\over\epsilon^2}\, \left({\mu^2\over |k_{\perp}|^2}\right)^\epsilon
\left[1 + \epsilon \psi(1-\epsilon) - \epsilon \psi(1+\epsilon)\right]\,
\Biggr]\,-\,{\beta_0\over2\epsilon}\Biggr\}\ .\nonumber
\end{eqnarray}
Eq.~(\ref{nllall}) is valid to all orders in $\epsilon$, and it does
not depend on $\delta_R$ or $N_f$ (except through the $\overline{\rm MS}$ 
ultraviolet counterterm).  Expanded to $O(\epsilon)$ 
it reads,
\begin{eqnarray}
\left.{{\rm Disp}\, C^{g(1)}_{\nu}(q_a,q_b)\over C^{g(0)}_{\nu}(q_a,q_b)}
\right|_{k\to 0} &&=\  
c_\Gamma\,\Biggl\{ N_c\, \Biggl[ {2\over\epsilon}\, \left({\mu^2\over -t}
\right)^\epsilon\,\ln{\tau\over |k_{\perp}|^2}  
  \nonumber\\ &&\qquad-
\left({\mu^2\over |k_{\perp}|^2}\right)^\epsilon \left( {1\over\epsilon^2}\, 
- {\pi^2\over 3}\right)\Biggr]\,-\,{\beta_0\over2\epsilon} 
+ O(\epsilon^2)\Biggr\}\, ,\label{nllone}
\end{eqnarray}
which to $O(\epsilon^0)$ agrees with the soft limit of
eq.~(\ref{looplip}).
Matching eq.~(\ref{nllall}) or eq.~(\ref{nllone}) to the 
correction to the Lipatov vertex (\ref{looplip}), we obtain the
one-loop Lipatov vertex with soft corrections to all orders in $\epsilon$ or 
to $O(\epsilon)$, respectively. The matching
yields
\begin{eqnarray}
\lefteqn{ {{\rm Disp}\,C^{g(1)}_{\nu}(q_a,q_b)\over C^{g(0)}_{\nu}(q_a,q_b)} = 
c_\Gamma\, \Biggl\{N_c\, \Biggl[ {1\over\epsilon}\, 
\left({\mu^2\over -t_1}\right)^\epsilon \ln{\tau\over
|k_{\perp}|^2} + {1\over\epsilon}\,\left({\mu^2\over -t_2}\right)^\epsilon
\ln{\tau\over |k_{\perp}|^2} } \nonumber\\ &-&
\left({\mu^2\over |k_{\perp}|^2}\right)^\epsilon
\left( {1\over\epsilon^2}\, - {\pi^2\over 3}\right)
-{1\over 2} \ln^2{t_1\over t_2} \Biggr] -{\beta_0\over2\epsilon}
- {\beta_0\over 2} \left(t_1 + t_2 + 2 q_{a\perp}
q_{b\perp}^\ast\right) {L_0(t_1/t_2)\over
t_2} \label{alllip}\\ &+& {N_c-N_f\over 3}\, |k_{\perp}|^2\,
\left[ - [ 2t_1 t_2 + (t_1 + t_2 + 2|k_{\perp}|^2) q_{a\perp} 
q_{b\perp}^\ast ] {L_2(t_1/t_2)\over t_2^3} - {q_{a\perp} 
q_{b\perp}^\ast \over 2\, t_1\, t_2} \right]
 \Biggr\} \, ,\nonumber
\end{eqnarray}
where this equation is exact to $O(\epsilon)$ in the $k\rightarrow0$
limit and to $O(\epsilon^0)$ elsewhere. Eq.~(\ref{alllip})
agrees with the NLL corrections to the Lipatov vertex computed
in ref.~\cite{linear,fadin} in the CDR scheme\footnote{
In ref.~\cite{linear,fadin} the 
$\overline{\rm MS}$ subtraction is defined with a
factor of $\Gamma(1+\epsilon)/(4\pi)^{2-\epsilon}$ as opposed to the full 
$c_\Gamma$
factor, eq.~(\ref{cgam}), chosen by us. This yields an accountable
difference at $O(\epsilon)$ between the NLL corrections to the Lipatov 
vertex of ref.~\cite{linear,fadin} and eq.~(\ref{alllip}). However, it should
not make any difference in the complete NLL corrections to the BFKL 
resummation as long as the same definition of 
$\overline{\rm MS}$ is used in both the virtual and real contributions.  This
is because both the real and virtual IR singularities will be multiplied
by the same overall factor, so that after their cancellation the finite 
remainder will be the same regardless
of what the $O(\epsilon)$ piece of the UV counterterm is. This argument holds
as long as one doesn't introduce the two-loop counterterm, which
appears at NNLL.}, to $O(\epsilon)$.

\section{Discussion and Conclusions}
\label{sec:disc}

In this paper we have computed the one-loop
corrections to the Lipatov vertex, eq.(\ref{looplip}), to
$O(\epsilon^0)$ in both the CDR and the dimensional reduction schemes.
This was done by expanding the complete one-loop five-gluon helicity 
amplitudes in the multi-Regge kinematics.
The result (\ref{looplip}) was then augmented by the one-loop corrections 
to $O(\epsilon)$ in the soft limit for the intermediate gluon, which is 
necessary due to the infrared divergence which arises in
the squared amplitude in this region of phase space.
In eq.~(\ref{nllall}) we  
computed the soft corrections to all orders in $\epsilon$, and
in eq.~(\ref{alllip}) we performed the matching to $O(\epsilon)$ 
with the one-loop Lipatov vertex (\ref{looplip}). Our final results are
independent of the RS parameter $\delta_R$ (\ref{cp}), and agree with
the one-loop Lipatov vertex computed in the CDR scheme to $O(\epsilon^0)$
in ref.~\cite{fl,virtual}, and to $O(\epsilon)$ in ref.~\cite{linear}.

The one-loop Lipatov vertex (\ref{alllip}), however, depends on
the arbitrary transverse scale $\tau$, introduced by the reggeization,
eq.~(\ref{sud}).  This dependence vanishes exactly if
we expand the high-energy $n$-gluon amplitudes to one-loop\footnote{
We have verified the logarithmic terms in this expansion for six gluons by
comparing with the maximal-helicity violating $N=4$ supersymmetric 
six-gluon amplitude at one-loop \cite{bddk}.},
using the reggeized gluon (\ref{sud}), the NLL impact factors
(\ref{allvert}), and the NLL Lipatov vertex (\ref{alllip})
as we did for the five-gluon amplitude (\ref{pres}).  
However, in the higher-loop contributions at NLL accuracy there will 
be residual dependence on the scale $\tau$.    
In particular, the terms proportional to $\ln(\tau/|k_\perp|^2)$ in
eq.~(\ref{alllip}), which cancel in the expansion of the amplitudes to
one loop, resurface again at two loops as double logarithms.
Although formally next-to-next-to-leading-logarithmic (NNLL), these terms
would ruin the subtle infrared cancellation between real and virtual gluon
integrations when $k_\perp$ becomes soft, thus invalidating the entire
BFKL program. 

The solution to this problem is suggested by the form of the high-energy
one-loop five-gluon amplitude (\ref{oneloop}).  In this equation we see
that the high-energy logarithms occur naturally in terms of the physical
rapidity intervals.  Therefore, it is reasonable to implement a
modified prescription for reggeizing the gluons. This prescription is to
replace
\begin{equation}
{1\over t_i} \to {1\over t_i} 
\left({s_{i-1,i}\over |k_{i-1\perp}||k_{i\perp}|}\right)^{\alpha(t_i)}
={1\over t_i} 
e^{\alpha(t_i)y_{i-1,i}}
\label{sud2}
\end{equation}
for each of the $t$-channel gluons, where $y_{i-1,i}$ is the physical
rapidity interval between the emitted gluons $i-1$ and $i$.
Effectively, this means that the scale $\tau$ 
``runs'' as we travel down the reggeon ladder, so that $\tau_i=
|k_{i-1\perp}||k_{i\perp}|$.  The terms proportional to 
$\ln(\tau/|k_\perp|^2)$  and $\ln(\tau/-t)$ in eqs.~(\ref{alllip}) and
(\ref{allvert}), respectively, vanish in this prescription so that these
potentially troublesome logarithms have been removed at NLL 
accuracy\footnote{An alternative to this prescription is to let $\tau_i=
\lambda |k_{i-1\perp}||k_{i\perp}|$, where $\lambda$ is some arbitrary
constant.  Then, the logarithms
$\ln(\tau/|k_\perp|^2)$  and $\ln(\tau/-t)$ each get replaced by
$\ln{\lambda}$.  The troublesome dependence on the $k_{\perp}$ of the
emitted gluon is still removed, but now one can test the sensitivity
of the NLL resummation to the arbitrary constant $\lambda$.  To be 
consistent, this prescription also requires the introduction of a 
corresponding $\lambda$-dependence in the real-gluon contributions.}.  
Although not stated explicitly in this language, the above
prescription is in effect used by Fadin and collaborators when they
change the scale of the one-loop Lipatov vertex and the one-loop 
impact factors to remove the dependence on $\tau$ to NLL \cite{fadin}.

\appendix
\section{Spinor Algebra in the Multi-Regge kinematics}
\label{sec:appa}

We consider the scattering of two gluons of momenta $p_a$ and $p_b$
into $n+2$ gluons of momenta $p_{i}$, where $i=a',b',1\dots n$.
 Using light-cone coordinates $p^{\pm}= p_0\pm p_z$, and
complex transverse coordinates $p_{\perp} = p_x + i p_y$, with scalar
product $2 p\cdot q = p^+q^- + p^-q^+ - p_{\perp} q^*_{\perp} - p^*_{\perp} 
q_{\perp}$, the gluon 4-momenta are,
\begin{eqnarray}
p_a &=& \left(p_a^+, 0; 0, 0\right)\, ,\nonumber \\
p_b &=& \left(0, p_b^-; 0, 0\right)\, ,\label{in}\\
p_i &=& \left(|p_{i\perp}| e^{y_i}, |p_{i\perp}| e^{-y_i}; 
|p_{i\perp}|\cos{\phi_i}, |p_{i\perp}|\sin{\phi_i}\right)\, ,\nonumber
\end{eqnarray}
where to the left of the semicolon we have the + and -
components, and to the right the transverse components.
$y$ is the gluon rapidity and $\phi$ is the azimuthal angle between the 
vector $p_{\perp}$ and an arbitrary vector in the transverse plane.
Momentum conservation gives
\begin{eqnarray}
0 &=& \sum p_{i\perp}\, ,\nonumber \\
p_a^+ &=& \sum p_{i}^+ \, ,\label{kin}\\ 
p_b^- &=& \sum p_{i}^- \, .\nonumber
\end{eqnarray}

For each massless momentum $p$ there is a positive and negative helicity 
spinor, $|p+\rangle$ and $|p-\rangle$, so we can consider two types of 
spinor products 
\begin{eqnarray}
        \langle pq\rangle & = & \langle p-|q+\rangle\nonumber\\
        \left[ pq \right] & = & \langle p+|q-\rangle\ .
        \label{spinors}
\end{eqnarray}
Phases are chosen so that $\langle pq\rangle=-\langle qp\rangle$
and $[pq]=-[qp]$.
For the momentum under consideration the spinor products are
\begin{eqnarray}
\langle p_{i} p_{j}\rangle &=& p_{i\perp}\sqrt{p_{j}^+
\over p_{i}^+} - p_{j\perp} \sqrt{p_{i}^+\over p_{j}^+}\, 
,\nonumber\\ \langle p_a p_i\rangle &=& -\sqrt{p_a^+
\over p_i^+}\, p_{i\perp}\, ,\label{spro}\\ \langle p_i p_b\rangle &=&
-\sqrt{p_i^+ p_b^-}\, ,\nonumber\\ \langle p_a p_b\rangle 
&=& - \sqrt{p_a^+ p_b^-} = -\sqrt{s_{ab}}\, ,\nonumber
\end{eqnarray}
where we have used the mass-shell condition 
$|p_{i\perp}|^2 = p_i^+ p_i^-$.  The other type of spinor product 
can be obtained from
\begin{equation}
        [pq]\ =\ \pm\langle qp\rangle^{*}\ ,
\end{equation}
where the $+$ is used if $p$ and $q$ are both ingoing or both 
outgoing, and the $-$ is used if one is ingoing and the other 
outgoing.

In the multi-Regge kinematics, the gluons are strongly ordered in 
rapidity and have comparable transverse momentum:
\begin{equation}
y_{a'} \gg y_1\gg\dots y_{n} \gg y_{b'};\qquad |p_{i\perp}|\simeq|p_{\perp}|\, 
.\nonumber
\end{equation}
Then the momentum conservation (\ref{kin}) in the $\pm$ directions reduces to
\begin{eqnarray}
p_a^+ &\simeq& p_{a'}^+\, ,\nonumber\\ 
p_b^- &\simeq& p_{b'}^-\, ,\label{hkin}
\end{eqnarray}
and the Mandelstam invariants become
\begin{eqnarray}
s_{ab} &=& 2 p_a\cdot p_b \simeq p_{a'}^+ p_{b'}^- \nonumber\\ 
s_{ai} &=& -2 p_a\cdot p_i \simeq - p_{a'}^+ p_i^- \label{mrinv}\\ 
s_{bi} &=& -2 p_b\cdot p_i \simeq - p_i^+ p_{b'}^- \nonumber\\ 
s_{ij} &=& 2 p_i\cdot p_j \simeq |p_{i\perp}| |p_{j\perp}| e^{y_i-y_j}
=p_{i}^{+}p_{j}^{-}\qquad (y_{i}\gg y_{j})\,
,\nonumber
\end{eqnarray}
where $i,j=a',b',1\dots n$.  
In this limit the
 spinor products (\ref{spro}) become
\begin{eqnarray}
\langle p_a p_b\rangle &\simeq& \langle p_{a'} p_b\rangle \simeq
-\sqrt{p_{a'}^+\over p_{b'}^+} |p_{b'\perp}| \nonumber\\
\langle p_a p_{b'}\rangle &\simeq& \langle p_{a'} p_{b'}\rangle =
-\sqrt{p_{a'}^+\over p_{b'}^+}\, p_{b'\perp} \nonumber\\
\langle p_a p_{a'}\rangle &\simeq& - p_{a'\perp} \nonumber\\
\langle p_{b'} p_b\rangle &\simeq& - |p_{b'\perp}| \label{hpro}\\
\langle p_a p_i\rangle &\simeq& \langle p_{a'} p_i\rangle =
-\sqrt{p_{a'}^+\over p_i^+}\, p_{i\perp} \nonumber\\
\langle p_i p_b\rangle &\simeq& -\sqrt{p_i^+\over p_{b'}^+} 
|p_{b'\perp}| \nonumber\\ 
\langle p_i p_{b'}\rangle &\simeq& -\sqrt{p_i^+\over p_{b'}^+} 
p_{b'\perp} \nonumber\\ 
\langle p_i p_{j}\rangle &\simeq& -\sqrt{p_i^+\over p_{j}^+}\,
p_{j\perp} \qquad (y_{i}\gg y_{j})\ .\nonumber
\end{eqnarray}

\section{Logarithmic Functions}
\label{sec:appb}

\begin{eqnarray}
L_0(x) &=& {\ln(x)\over 1-x}\nonumber\\
L_1(x) &=& {\ln(x) +1-x\over (1-x)^2} \nonumber\\
L_2(x) &=& {1\over (1-x)^3} \left[\ln(x) -{x\over 2} +{1\over 2x} \right] 
\label{logar}\\ Ls_1(x,y) &=& {1\over (1-x-y)^2}\, \left[ {\rm Li}_2(1-x) 
+ {\rm Li}_2(1-y) \right. \nonumber\\ && \left. + \ln(x)\,\ln(y) + 
(1-x-y) [ L_0(x)+L_0(y) ] - {\pi^2\over 6} \right]\, ,\nonumber
\end{eqnarray}
where ${\rm Li}_2$ is the dilogarithm.

\section{The Absorptive Parts}
\label{sec:appc}

In the subamplitudes of type $m_{5:1}$
the absorptive parts arise entirely from the universal 
piece~(\ref{vertica})
\begin{eqnarray}
\lefteqn{ {\rm Absorp}\, m_{5:1}(A^-,A'^+,k^+,B'^+,B^-)}
\nonumber\\ &&=\  
- m_5(A^-,A'^+,k^+,B'^+,B^-)\, c_{\Gamma}\, N_c\, {\pi\over\epsilon}\,
\left[ \left({\mu^2\over -t_1}\right)^\epsilon +
\left({\mu^2\over |k_{\perp}|^2} \right)^\epsilon +
\left({\mu^2\over -t_2}\right)^\epsilon  \right] \nonumber\\
\lefteqn{ {\rm Absorp}\, m_{5:1}(A^-,A'^+,k^+,B^-,B'^+)}
\nonumber\\ &&=\  
- m_5(A^-,A'^+,k^+,B^-,B'^+)\, c_{\Gamma}\, N_c\, {\pi\over\epsilon}\,
\left[ \left({\mu^2\over -t_1}\right)^\epsilon +
\left({\mu^2\over |k_{\perp}|^2} \right)^\epsilon -
\left({\mu^2\over -t_2}\right)^\epsilon  \right] 
\label{absorp}\\
\lefteqn{ {\rm Absorp}\, m_{5:1}(A^-,A'^+,B'^+,B^-,k^+)}
\nonumber\\ &&=\  
- m_5(A^-,A'^+,B'^+,B^-,k^+)\, c_{\Gamma}\, N_c\, {\pi\over\epsilon}\,
\left[ \left({\mu^2\over -t_1}\right)^\epsilon -
\left({\mu^2\over |k_{\perp}|^2} \right)^\epsilon +
\left({\mu^2\over -t_2}\right)^\epsilon  \right] \nonumber\\
\lefteqn{ {\rm Absorp}\, m_{5:1}(A^-,k^+,B'^+,B^-,A'^+)}
\nonumber\\ &&=\  
- m_5(A^-,k^+,B'^+,B^-,A'^+)\, c_{\Gamma}\, N_c\, {\pi\over\epsilon}\,
\left[ - \left({\mu^2\over -t_1}\right)^\epsilon +
\left({\mu^2\over |k_{\perp}|^2} \right)^\epsilon +
\left({\mu^2\over -t_2}\right)^\epsilon  \right]\, . \nonumber
\end{eqnarray}

The subamplitudes of type $m_{5:3}$ also contribute 
to the absorptive part. Because of the reflection and
cyclic symmetries of these subamplitudes~\cite{bk1},
\begin{equation}
m_{5:3}(1,2;3,5,4) = m_{5:3}(2,1;5,4,3) = - m_{5:3}(1,2;3,4,5)\, 
,\label{refl}
\end{equation}
we may rewrite their contribution 
to the one-loop five-gluon amplitude~(\ref{loop}) as
\begin{eqnarray}
\lefteqn{ \sum_{S_5/Z_2\times Z_3} {\rm tr}(\lambda^{d_{\sigma(1)}} 
\lambda^{d_{\sigma(2)}}) {\rm tr}(\lambda^{d_{\sigma(3)}} 
\lambda^{d_{\sigma(4)}}\lambda^{d_{\sigma(5)}})\, m_{5:3}(\sigma(1), 
\sigma(2); \sigma(3), \sigma(4), \sigma(5)) } \nonumber\\ &=&
{i\over 4}\, \sum_{S_5/Z_2\times P_3}
\delta^{d_{\sigma(1)} d_{\sigma(2)}}\, f^{d_{\sigma(3)}
d_{\sigma(4)} d_{\sigma(5)}}\, m_{5:3}(\sigma(1), 
\sigma(2); \sigma(3), \sigma(4), \sigma(5))\, ,\label{permut}
\end{eqnarray}
where the sums are over the permutations of the five color indices, 
up to cyclic permutations within each trace on the left-hand side and
up to any permutation within each trace on the right-hand side. This 
singles out ten different subamplitudes of type $m_{5:3}$ to be
computed.  Using the equation for the decomposition of 
$m_{5:3}$ subamplitudes in terms of
$m_{5:1}$ subamplitudes, eq.~(\ref{nonp}),
%
%
and taking into account the eight leading color orderings of the
subamplitudes of type $m_{5:1}$ elucidated in sect.~\ref{sec:provb},
we find that at NLL there are only three different values for the 
subamplitudes of type $m_{5:3}$.  They satisfy the following relations 
\begin{eqnarray}
\lefteqn{ m_{5:3}(A^-,B^-;A'^+,k^+,B'^+) = m_{5:3}(A'^+,B^-;A^-,k^+,B'^+)}
\nonumber\\ &&=\ m_{5:3}(A^-,B'^+;A'^+,k^+,B^-) = 
m_{5:3}(A'^+,B'^+;A^-,k^+,B^-)\, ;\nonumber\\ 
\lefteqn{ m_{5:3}(A^-,A'^+;k^+,B'^+,B^-) } \label{explnll}\\ &&=\  
m_{5:3}(A^-,k^+;A'^+,B'^+,B^-) = m_{5:3}(A'^+,k^+;A^-,B'^+,B^-)\, ;
\nonumber\\ \lefteqn{ m_{5:3}(B^-,B'^+;A^-,A'^+,k^+) } \nonumber\\ &&=\  
m_{5:3}(B^-,k^+;A^-,A'^+,B'^+) = m_{5:3}(B'^+,k^+;A^-,A'^+,B^-)\, .\nonumber
\end{eqnarray}
Using eq.~(\ref{nonp},\ref{vg}, \ref{nlla}) one can easily
check that the dispersive parts of the $m_{5:3}$ subamplitudes 
cancel, as expected.  The absorptive parts are, using eq.(\ref{absorp}), 
\begin{eqnarray}
\lefteqn{ {\rm Absorp}\, m_{5:3}(A^-,B^-;A'^+,k^+,B'^+) } \nonumber\\ &&=\  
- m_5(A^-,A'^+,k^+,B'^+,B^-)\, c_{\Gamma}\, {2\pi\over\epsilon}\,
\left[ \left({\mu^2\over -t_1}\right)^\epsilon -
\left({\mu^2\over |k_{\perp}|^2} \right)^\epsilon +
\left({\mu^2\over -t_2}\right)^\epsilon  \right] \nonumber\\
\lefteqn{ {\rm Absorp}\, m_{5:3}(A^-,A'^+;k^+,B'^+,B^-) } \label{trenll}\\
&&=\ - m_5(A^-,A'^+,k^+,B'^+,B^-)\, c_{\Gamma}\, {2\pi\over\epsilon}\,
\left[ \left({\mu^2\over -t_1}\right)^\epsilon +
\left({\mu^2\over |k_{\perp}|^2} \right)^\epsilon -
\left({\mu^2\over -t_2}\right)^\epsilon  \right] \nonumber\\
\lefteqn{ {\rm Absorp}\, m_{5:3}(B^-,B'^+;A^-,A'^+,k^+) } \nonumber\\ 
&&=\ - m_5(A^-,A'^+,k^+,B'^+,B^-)\, c_{\Gamma}\, {2\pi\over\epsilon}\,
\left[ -\left({\mu^2\over -t_1}\right)^\epsilon +
\left({\mu^2\over |k_{\perp}|^2} \right)^\epsilon +
\left({\mu^2\over -t_2}\right)^\epsilon  \right]\, .\nonumber
\end{eqnarray}

Finally, we can combine these to find
\begin{eqnarray}
\lefteqn{ {\rm Absorp}\, M_{5}^{1-loop}(A^-,A'^+,k^+,B'^+,B^-) 
\ =\  g^5\, c_\Gamma\,N_c\,{\pi\over\epsilon}}\nonumber \\
&&\qquad
\times\ 2 {s}\, C_{-\nu_a\nu_{a'}}^{gg(0)}(-p_a,p_{a'})\,{1\over t_1}\, 
C^{g(0)}_{\nu}(q_a,q_b)\, {1\over t_2}\, 
C_{-\nu_b\nu_{b'}}^{gg(0)}(-p_b,p_{b'})\,\Biggl\{\nonumber\\ 
&&\qquad\qquad 
\left[ \left({\mu^2\over -t_1}\right)^\epsilon +
\left({\mu^2\over |k_{\perp}|^2} \right)^\epsilon +
\left({\mu^2\over -t_2}\right)^\epsilon  \right] \,
{1\over 2} if^{aa'c}if^{cdc'}if^{c'b'b}\nonumber\\
&&\qquad\quad 
+\left[ \left({\mu^2\over -t_1}\right)^\epsilon -
\left({\mu^2\over |k_{\perp}|^2} \right)^\epsilon +
\left({\mu^2\over -t_2}\right)^\epsilon  \right] \,
\left[{i\over2}d^{aa'c}f^{cdc'}d^{c'b'b} \right.\nonumber\\
&&\qquad\qquad\qquad\qquad\left.
+{i\over N_c}\left(\delta^{ab}f^{a'db'}+
\delta^{a'b}f^{adb'}+\delta^{ab'}f^{a'db}+
\delta^{a'b'}f^{adb}\right)\right]
\\ &&\qquad\quad 
+\left[ \left({\mu^2\over -t_1}\right)^\epsilon +
\left({\mu^2\over |k_{\perp}|^2} \right)^\epsilon -
\left({\mu^2\over -t_2}\right)^\epsilon  \right] \,
\left[{i\over2}d^{aa'c}d^{cdc'}f^{c'b'b}
\right.\nonumber\\
&&\qquad\qquad\qquad\qquad\left.
+{i\over N_c}\left(2\delta^{aa'}f^{db'b}+
\delta^{ad}f^{a'b'b}+\delta^{a'd}f^{ab'b}\right)\right]
\nonumber\\
&&\quad \qquad
+\left[- \left({\mu^2\over -t_1}\right)^\epsilon +
\left({\mu^2\over |k_{\perp}|^2} \right)^\epsilon +
\left({\mu^2\over -t_2}\right)^\epsilon  \right] \,
\left[{i\over2}f^{aa'c}d^{cdc'}d^{c'b'b}
\right.\nonumber\\
&&\qquad\qquad\qquad\qquad\left.\left.
+{i\over N_c}\left(2\delta^{bb'}f^{aa'd}+
\delta^{bd}f^{aa'b'}+\delta^{b'd}f^{aa'b}\right)\right]\right\}\ 
.\nonumber
\end{eqnarray}

\vspace{.5cm}

{\sl Acknowledgements}
This work was partly supported by the
NATO Collaborative Research Grant CRG-950176, and
by the EU Fourth Framework Programme `Training and
Mobility of Researchers', Network `Quantum Chromodynamics and the 
Deep Structure of
Elementary Particles', contract FMRX-CT98-0194 (DG 12 - MIHT).

\end{document}